\documentclass[reprint,superscriptaddress,nofootinbib,amsmath,amssymb,aps,prl]{revtex4-2}
\usepackage[T1]{fontenc}
\usepackage{amsmath}
\usepackage{xcolor}
\usepackage{physics}
\usepackage{graphicx}% Include figure files
\usepackage{dcolumn}% Align table columns on decimal point
\usepackage{bm}% bold math
\usepackage{hyperref}% add hypertext capabilities
\usepackage{array}
\usepackage{ulem}% add strike out command \sout
\usepackage{etoolbox}
\usepackage{graphicx}
\DeclareUnicodeCharacter{00A0}{}
\usepackage{lipsum}
\usepackage{xr}
\usepackage{subfiles}

\emergencystretch=\maxdimen
\allowdisplaybreaks

\makeatletter
\patchcmd{\@makecaption}
  {\scshape}
  {}
  {}
  {}
\makeatother

\usepackage{xr}
\begin{document}

% \preprint{APS/123-QED}

\title{Enhancement of Electric Drive in Silicon Quantum Dots with Electric Quadrupole Spin Resonance}

\author{Philip Y. Mai}
    \thanks{These authors contributed equally}
    \affiliation{
        School of Electrical Engineering and Telecommunications, \\
        The University of New South Wales, Sydney, NSW 2052, Australia
    }
    \affiliation{
        Diraq, Sydney, NSW 2052, Australia
    }

\author{Pedro H. Pereira}
    \thanks{These authors contributed equally}
    \affiliation{
        Instituto de Física, Universidade Federal Fluminense, 24210-346, RJ, Brazil
    }

\author{Lucas Andrade Alonso}
    \thanks{These authors contributed equally}
    \affiliation{
        Instituto de Física, Universidade Federal Fluminense, 24210-346, RJ, Brazil
    }

\author{Ross C. C. Leon}
    \thanks{Current Address: Quantum Motion, 9 Sterling Way, London N7 9HJ, United Kingdom}
    \affiliation{
        School of Electrical Engineering and Telecommunications, \\
        The University of New South Wales, Sydney, NSW 2052, Australia
    }
\author{Chih Hwan Yang}
    \affiliation{
        School of Electrical Engineering and Telecommunications, \\
        The University of New South Wales, Sydney, NSW 2052, Australia
    }
    \affiliation{
        Diraq, Sydney, NSW 2052, Australia
    }
\author{Jason C. C. Hwang}
    \affiliation{
        School of Electrical Engineering and Telecommunications, \\
        The University of New South Wales, Sydney, NSW 2052, Australia
    }
\author{Daniel Dunmore}
    \affiliation{
        School of Electrical Engineering and Telecommunications, \\
        The University of New South Wales, Sydney, NSW 2052, Australia
    }
\author{Julien Camirand Lemyre}
    \affiliation{
        Institut Quantique et Département de Physique, Université de Sherbrooke, Sherbrooke, Québec, J1K 2R1, Canada
    }
\author{Tuomo Tanttu}
    \affiliation{
        School of Electrical Engineering and Telecommunications, \\
        The University of New South Wales, Sydney, NSW 2052, Australia
    }
    \affiliation{
        Diraq, Sydney, NSW 2052, Australia
    }
\author{Wister Huang}
    \thanks{
        Current Address: Laboratory for Solid State Physics, ETH Zurich, CH-8093 Zurich, Switzerland
    }
    \affiliation{
        School of Electrical Engineering and Telecommunications, \\
        The University of New South Wales, Sydney, NSW 2052, Australia
    }
\author{Kok Wai Chan}
    \affiliation{
        School of Electrical Engineering and Telecommunications, \\
        The University of New South Wales, Sydney, NSW 2052, Australia
    }
    \affiliation{
        Diraq, Sydney, NSW 2052, Australia
    }
\author{Kuan Yen Tan}
    \thanks{
        Current Address: IQM Quantum Computers, Espoo 02150, Finland
    }
    \affiliation{
        School of Electrical Engineering and Telecommunications, \\
        The University of New South Wales, Sydney, NSW 2052, Australia
    }
\author{Jesús D. Cifuentes}
    \affiliation{
        School of Electrical Engineering and Telecommunications, \\
        The University of New South Wales, Sydney, NSW 2052, Australia
    }
    \affiliation{
        Diraq, Sydney, NSW 2052, Australia
    }
\author{Fay E. Hudson}
    \affiliation{
        School of Electrical Engineering and Telecommunications, \\
        The University of New South Wales, Sydney, NSW 2052, Australia
    }
    \affiliation{
        Diraq, Sydney, NSW 2052, Australia
    }
\author{Kohei M. Itoh}
    \affiliation{
        School of Fundamental Science and Technology, Keio University, 3-14-1 Hiyoshi, Kohokuku, Yokohama, 223-8522, Japan
    }
\author{Arne Laucht}
    \affiliation{
        School of Electrical Engineering and Telecommunications, \\
        The University of New South Wales, Sydney, NSW 2052, Australia
    }
    \affiliation{
        Diraq, Sydney, NSW 2052, Australia
    }
\author{Michel Pioro-Ladrière}
    \affiliation{
        Institut Quantique et Département de Physique, Université de Sherbrooke, Sherbrooke, Québec, J1K 2R1, Canada
    }
    \affiliation{
        Quantum Information Science Program, Canadian Institute for Advanced Research, Toronto, ON, M5G 1Z8, Canada
    }
\author{Christopher C. Escott}
    \affiliation{
        School of Electrical Engineering and Telecommunications, \\
        The University of New South Wales, Sydney, NSW 2052, Australia
    }
    \affiliation{
        Diraq, Sydney, NSW 2052, Australia
    }
\author{Andrew Dzurak}
    \affiliation{
        School of Electrical Engineering and Telecommunications, \\
        The University of New South Wales, Sydney, NSW 2052, Australia
    }
    \affiliation{
        Diraq, Sydney, NSW 2052, Australia
    }
\author{Andre Saraiva}
    %\email{a.saraiva@unsw.edu.au}
    \affiliation{
        School of Electrical Engineering and Telecommunications, \\
        The University of New South Wales, Sydney, NSW 2052, Australia
    }
    \affiliation{
        Diraq, Sydney, NSW 2052, Australia
    }
\author{Reinaldo de Melo e Souza}
    \affiliation{
        Instituto de Física, Universidade Federal Fluminense, 24210-346, RJ, Brazil
    }
\author{MengKe Feng}
    \email{mengke.feng@unsw.edu.au}
    \affiliation{
        School of Electrical Engineering and Telecommunications, \\
        The University of New South Wales, Sydney, NSW 2052, Australia
    }
    \affiliation{
        Diraq, Sydney, NSW 2052, Australia
    }

\begin{abstract}
Quantum computation with electron spin qubits requires coherent and efficient manipulation of these spins, typically accomplished through the application of alternating magnetic or electric fields for electron spin resonance (ESR). In particular, electrical driving allows us to apply localized fields on the electrons, which benefits scale-up architectures. However, we have found that Electric Dipole Spin Resonance (EDSR) is insufficient for modeling the Rabi behavior in recent experimental studies. Therefore, we propose that the electron spin is being driven by a new method of electric spin qubit control which generalizes the spin dynamics by taking into account a quadrupolar contribution of the quantum dot: electric quadrupole spin resonance (EQSR). In this work, we explore the electric quadrupole driving of a quantum dot in silicon, specifically examining the cases of 5 and 13 electron occupancies.
\end{abstract}

\maketitle

\section{Introduction}

    %Qubits, the fundamental building block of quantum computers, work by storing quantum information in two quantum states. 
    %Therefore, implementations of a quantum computer are possible as long as there are two quantum states (for each qubit) that can be coherently controlled and manipulated to interact with other qubits. 
    Qubits are the fundamental building blocks of quantum computers and must be coherently controlled and manipulated to interact with other qubits.
    Such implementations include electron spin qubits implemented using silicon Metal-Oxide Semiconductor (MOS) technology, which offer numerous advantages. High-fidelity quantum operations have been demonstrated in silicon \cite{Yoneda2017,Huang2019,mkadzik2022precision,noiri2022fast,xue2022quantum,tanttu2024assessment} but the main advantage over other qubit technologies is the ability to leverage existing MOS technology. That provides the possibility of creating a universal quantum computer that could solve problems otherwise too costly for classical computers \cite{Veldhorst2017,saraiva2022materials}. The large number of qubits allows quantum error correction algorithms to be implemented, protecting fragile qubits from external sources of noise that can destroy quantum information \cite{Devitt2013}. 
    
    Electron spins can be coherently controlled using a microwave source to generate an AC magnetic field resonant with the electronic energy level splitting due to the Zeeman effect \cite{Veldhorst2017}. However, using the magnetic field for individual qubit addressability can be challenging because the field is not localized in space. % There are current research efforts to address this problem by using an ``always on'' global magnetic field \cite{Laucht2015,vahapoglu2021single, vahapoglu2022coherent} such as the recently proposed SMART protocol \cite{hansen2021pulse,seedhouse2021quantum,hansen2022implementation,hansen2024entangling}. 
    
    Alternatively, electric driving is localized 
    %offers another solution to this problem because electric fields are much more localized. 
    but the electron requires a position and/or momentum-dependent magnetic field to couple the spin degree of freedom to the electric field. This spin-orbit coupling (SOC) %is commonly called spin orbit coupling (SOC) and 
    can be realized with an external micro-magnet \cite{obata2010coherent,takeda2016fault,huang2021fast,kawakami2014electrical}, or with intrinsic SOC in silicon MOS heterostructures \cite{Tanttu2019,gilbert2023demand}, enabling Rabi frequencies up to tens of megahertz \cite{Yoneda2017}, which can be an order of magnitude faster compared to the typical magnetic electron spin resonance (ESR) drive \cite{Pla2012}. % There have also been novel schemes of electric driving to enhance its effect \cite{teske2023flopping}, as well as studies of achieving high-fidelity control \cite{rimbach2023simple}. 
    % Moreover, a key advantage of using intrinsic SOC over the micromagnet is the reduction of engineering challenges in scale up \cite{Veldhorst2017}. 
    The most common way to electrically drive a qubit is called Electric Dipole Spin Resonance (EDSR), where Rabi frequencies are promoted by an alternating electric field coupled with the electron orbital degree of freedom in the dipolar approximation.
    
    In a demonstration of electrically driven multielectron spin qubits using MOS quantum dots \cite{leon2020coherent}, it was shown that a multielectron dot not only mitigates charge noise and disorder \cite{Barnes2011}, but also allows for an order of magnitude enhancement %\sout{in Rabi frequency compared to} 
    that does not occur for a single electron \cite{ercan2023multielectron}. In particular, with 5 and 13 electrons, a higher quality factor was observed. This is significant and we seek to understand and reliably reproduce this behavior in a scale-up architecture. % This is a significant step towards having fault-tolerant qubits, and therefore, we seek to understand and reliably reproduce this behavior in a scale-up architecture.
    
    In this paper, we develop a theoretical model based on the experimental results obtained in Ref.~\cite{leon2020coherent}. We begin with a model describing the electric driving of orbital states which controls the spin-orbital character of driven states via the electrostatic gates \cite{gilbert2023demand}. Then, we add to the model by arguing that the EDSR description is not sufficient to describe the observed enhancements in Rabi frequency of multielectron dots with > 3 electrons, and we propose a new method named ``Electric Quadrupole Spin Resonance'' (EQSR).

    %In the following sections, we first provide a model of the device reported in Ref.~\cite{leon2020coherent}, making explicit the states that form the spin-orbit qubit for the 5 electrons case and how that leads to EQSR coupling. Then, we present theoretical evidence for EQSR by comparing the theoretical results of this modeling with the experimental results. Finally, we conclude by discussing how we can leverage EQSR in the future.

\section{EQSR Theory and Model}
    
    \subsection{Microscopic Model}
    \label{section:model}

    In this section, we build the effective Hamiltonian that describes the system reported in Ref.~\cite{leon2020coherent}. We will model the dot with an effective single-particle 2D Hamiltonian 
    %that describes the low energy physics as follows:
    \begin{align}
        H_\mathrm{DC} = H_\mathrm{xy} + H_\mathrm{Zee} + H_\mathrm{V} + H_\mathrm{SO}\; .
        \label{eq:h_dc} 
    \end{align}
    The name DC refers to the static parts of the physics - later we will look at the AC driving terms. This simplified Hamiltonian contains the essential physics, for the complete one used in the fitting see\cite{supp}.
    %This is a simplified description of the system, that illustrates the core physics presented here. The complete Hamiltonian is shown in the Supplementary Material \cite{supp} with additional terms relevant to the system that are necessary for fitting but not for the EQSR description.
    
    Let us examine the different components that make up $H_\mathrm{DC}$. The orbital Hamiltonian (deduced from COMSOL electrostatic simulations) and the kinetic energy (from silicon band structure) in the xy plane are given by
    \begin{equation}
        H_\mathrm{xy} = \frac{1}{2m_\mathrm{t}}(p_x^2+p_y^2) + \frac{1}{2}m_\mathrm{t}\omega_y^2(\delta^2x^2+y^2) \;,
    \label{eq:h_orb} 
    \end{equation}
    where $m_\mathrm{t}$ is the transverse effective mass in the $\Delta$ valley of silicon, $\delta\equiv \omega_x / \omega_y$ describes the ellipticity of the parabolic confinement ($\delta=1$ corresponds to a perfectly circular dot) and $p_{x(y)}$ is the kinetic momentum in the x(y) direction. The spin degree of freedom enters via the Zeeman Hamiltonian $H_\mathrm{Zee} = \frac{g\mu_B}{2}\mathbf{B}_0\cdot\bm{\sigma}$, with $\mathbf{B}_0$ being a static uniform magnetic field. $H_\mathrm{V}$ describes the degeneracy split of two z-valley states, due to the Si/SiO$_2$ interface, with an energy splitting likely between $500$ $\mathrm{\mu eV}$ to $1$ meV \cite{leon2020coherent, yang2013spin}.
    %{\color{blue}$H_\mathrm{V}$ is the Hamiltonian that represents the degeneracy split of two z-valley states, due to the Si/SiO$_2$ interface, with an energy splitting of hundreds of $\mathrm{\mu eV}$ as measured in Ref. \cite{leon2020coherent}.} 
    These three components can be treated as the unperturbed part of the Hamiltonian, and their eigenstates written as $|n,s,v_\pm\rangle$, where $n \in \{S,P_x,P_y,\dots\}$ denotes the 2D quantum harmonic oscillator orbital, $s\in\{\uparrow, \downarrow \}$ denotes the spin state and $v_\pm$ denotes the two valley states. The valley degree of freedom will not be relevant in the qubit dynamics (see below), so from now on we omit it.
    % {\color{blue}As we will see, the valley degree of freedom will not be relevant in the qubit dynamics, so from now on we will omit $v_\pm$ when writing the eigenstates.}

    The spin-orbit component $H_\mathrm{SO}$ is considered as a perturbation to $H_\mathrm{xy} + H_\mathrm{Zee}$. It promotes a hybridization of orbital and spin degrees of freedom between the unperturbed eigenstates $|n,s\rangle$. Here we write
        \begin{eqnarray}
            H_\mathrm{SO} = \alpha\underbrace{(k_x\sigma_z'+k_y\sigma_x')}_{h_\mathrm{R}}
            + \beta\underbrace{(-k_x\sigma_z'+k_y\sigma_x')}_{h_\mathrm{D}} \; ,
        \label{eq:h_so}
        \end{eqnarray}
    where $h_\mathrm{R}, h_\mathrm{D}$ are the Rashba and Dresselhaus terms with respective coefficients $\alpha, \beta$ which arise due to the lowered symmetry of the heterostructure and Si/SiO$_2$ interface respectively \cite{Golub2004}. $\sigma'$ indicates the Pauli matrices in the spin coordinates \cite{supp}.

    \begin{figure}
        \centering
        \includegraphics[width=0.4\textwidth, angle = 0]{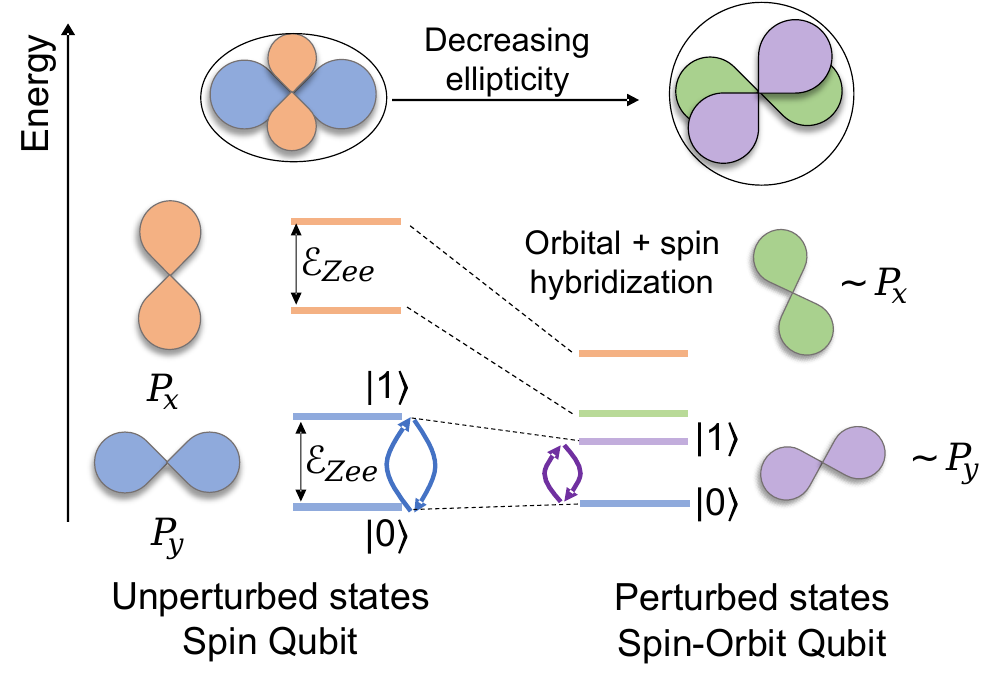}
        % \captionsetup{format=plain, font=small, labelfont=bf}
        \caption{\textbf{Schematic of p-orbital-like states.} 
       As the ellipticity of the electrostatic potential decreases, SOC hybridization of the orbital and spin characters of the states increases, forming a spin-orbit qubit. $\mathcal{E}_\text{Zee}$ is the Zeeman splitting.}
        \label{fig:hybridization}
    \end{figure}

    The experiments of Ref. \cite{leon2020coherent} were done with 5 and 13 electrons, which can be approximated within single-particle theory for the valence electron because 4 and 12 electrons are closed shell configurations. 

     %{\color{blue}The experiments of Ref. \cite{leon2020coherent} were done with a single electron outside the closed shells, justifying the single-particle theory approximation}.
    In the 5-electron experiment, the qubit is defined by the unperturbed states $|0\rangle_\mathrm{unp}=|P_y,\uparrow\rangle$ and $|1\rangle_\mathrm{unp}=|P_y,\downarrow\rangle$. However, $H_\mathrm{SO}$ couples the different orbital states, resulting in these states becoming
    \small
    \begin{equation}
    \begin{split}
        |0\rangle=\mathcal{N}_0(|{P_y,\uparrow}\rangle + a_0 |{S,\downarrow}\rangle + b_0 |{P_x,\downarrow}\rangle + \cdots)\sim|{P_y,\uparrow}\rangle \; ,\\
        |1\rangle=\mathcal{N}_1(|{P_y,\downarrow}\rangle + a_1 |{S,\uparrow}\rangle + b_1 |{P_x,\uparrow}\rangle + \cdots)\sim|{P_y,\downarrow}\rangle \; ,
    \label{eq:state_expansion}
    \end{split}
    \end{equation}
    \normalsize
    where $a_0$ comes from first order effect of $H_\mathrm{SO}$ (from the transition $|{P_y,\uparrow}\rangle\xrightarrow{k_y\sigma_x'}|{S,\downarrow}\rangle$), $b_0$ comes from second order effect of $H_\mathrm{SO}$ (from the transition $|{P_y,\uparrow}\rangle \xrightarrow{k_y\sigma_x'} |{S,\downarrow}\rangle \xrightarrow{k_x\sigma_z'} |{P_x,\downarrow}\rangle$), and $\mathcal{N}_0$ is a normalization constant. In a similar fashion the coefficients $a_1$, $b_1$ and $\mathcal{N}_1$ are defined. The approximation on the right hand side of the equation indicates that the original, unperturbed states are still the most dominant component. The states in Eq.~(\ref{eq:state_expansion}) differ both by spin and orbit degrees of freedom and the hybridization promoted by the spin-orbit Hamiltonian is increased when the dot ellipticity decreases (that is, when $\delta$ approaches 1), as analyzed in the next section. Fig. \ref{fig:hybridization} shows a schematic of this hybridization for p-orbital states.      
   The hybridization of spin and orbit degrees of freedom allows to drive the qubit electrically, while the valley,for our purposes, means only that we can place two electrons per $|n,s\rangle$ state (justifying its omission from now on).
    % {\color{blue}The hybridization of spin and orbit degrees of freedom is responsible for being possible to drive the qubit electrically, justifying the omission of the valley degree of freedom, which, for the scope of this paper, is relevant only for doubling the degeneracy of the states.}

    Based on Eq.~(\ref{eq:h_so}) and the coupling pathways of the states via $H_\mathrm{SO}$, we can determine that $a_{0(1)}\propto\alpha+\beta$ and $b_{0(1)}\propto\alpha^2-\beta^2$ \cite{supp}. These proportionalities lead to different dependencies of the Rabi speed-up on spin-orbit coupling. 
    % {\color{blue}The explicit perturbative analysis can be found in the Supplementary Material.}

    \subsection{Electric Quadrupole Spin Resonance (EQSR)}
     \label{section:EQSR}

    %Now that we have described the DC component of the Hamiltonian, we will 
    Let us now describe the AC driving component. An alternating electric field $\mathbf{E}=E_y\cos(\omega t)\hat{y}$ interacts with an electron according to the following Hamiltonian:
    \begin{equation}
        H_\mathrm{AC}^\mathrm{EDSR}(t)=|e|E_y y\cos(\omega t)\;,
    \end{equation}
    which allows spin-flip processes with the Rabi frequency $\Omega$ given by
    \begin{equation}
        \hbar\Omega_\mathrm{EDSR} = |\langle 0 | eE_y y| 1 \rangle |\;.\label{defRabiEDSR}
    \end{equation}
    This method of driving is called Electric Dipole Spin Resonance (EDSR). With 5 electrons present in the dot, the states $|0\rangle$ and $|1\rangle$ are those described in Eq.~(\ref{eq:state_expansion}), respectively. Therefore, the Rabi frequency would depend on the $a_{0(1)}$ coefficients, since the $\hat{y}$ operator connects states differing in one orbital excitation in the $y$ direction, in this case, the states $|P_y\rangle$ and $|S\rangle$. This means that a Rabi frequency enhancement would only occur when $a_{0}$ is large, i.e. when the $|P_y,\uparrow\rangle$ and $|S,\downarrow\rangle$ orbitals come close in energy: $\hbar\omega_y\sim \mathcal{E}_\text{Zee}$, enabling spin-orbit coupling to hybridize the two in a significant fraction ($\mathcal{E}_\text{Zee}$ is the Zeeman energy splitting). However, for typical values used in experiments, $\hbar\omega_y$ ($\sim10\mathrm{meV}$ \cite{Yang2012}) is many orders of magnitude larger than $\mathcal{E}_\text{Zee}$ ($\sim100\mathrm{\mu eV}$ \cite{leon2020coherent}). This indicates that EDSR cannot be responsible for the observed Rabi enhancement.

    \begin{figure}
        \centering
        \includegraphics[width=0.45\textwidth, angle = 0, trim = 0cm 3.5cm 12cm 0cm]{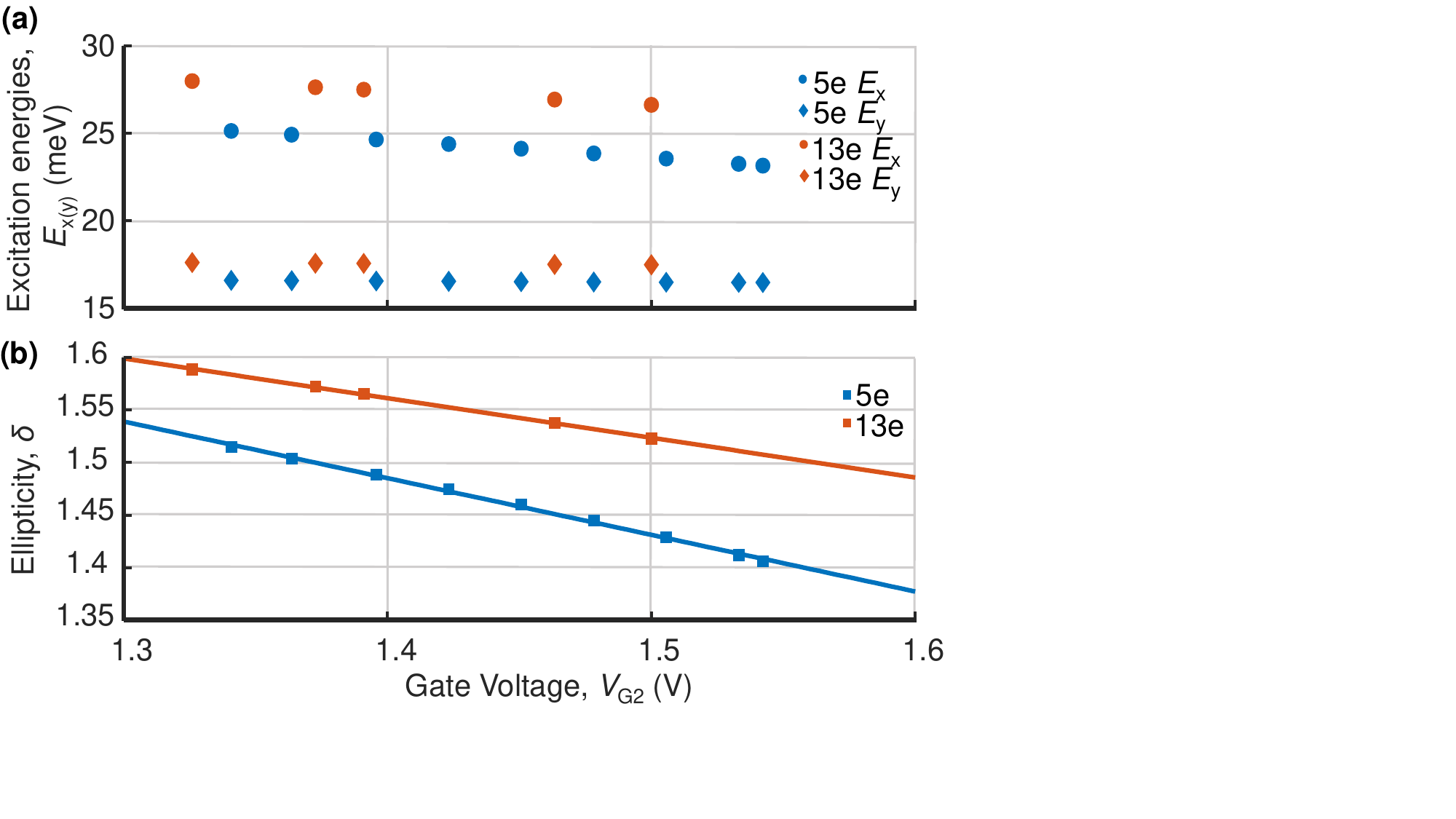}
        % \captionsetup{format=plain, font=small, labelfont=bf}
        \caption{\textbf{Ellipticity and orbital energies.} 
        (a) Plot of the extracted orbital excitation energies $\mathcal{E}_{x(y)}=\hbar\omega_{x(y)}$ against $V_\mathrm{G2}$ via real-space simulations of the single particle states given an electrostatic potential simulated with COMSOL. 
        (b) Plot of the ellipticity of the dot $\delta \equiv \frac{\mathcal{E}_{x}}{\mathcal{E}_{y}}$ against $V_\mathrm{G2}$ for the 5e configuration in Figs. 3(g-i) and 13e configuration in Figs. 3(j-l) of Ref.~\cite{leon2020coherent}. Linear slopes are also shown. Ellipticities were obtained from calculations of energies shown in (a).}
        \label{fig:plot_orbital_energies_1}
    \end{figure}
    
    Then, what can cause the enhancement? The experiments give us two clues. First, no enhancement is observed for a single electron dot \cite{leon2020coherent}, suggesting a possible dependence on a property of the $P$ or $D$ orbitals, not present in the $S$ orbitals. The main difference here is the 2- or 3-fold orbital degeneracy. 
    
    Second, as the multi-electron dot is tuned towards the degeneracy point (by increasing the electrostatic gate voltage $V_\mathrm{G2}$), the lateral electrostatic confinement is altered. We confirm this via real-space simulations of the COMSOL electrostatic simulations of the device at the operating point, extracting only the single-particle states from electronic structure simulation tools \cite{wang2023jellybean}. From the simulations, we obtained the excitation energies $\mathcal{E}_x=\hbar\omega_x$ (circles) and $\mathcal{E}_y=\hbar\omega_y$ (diamonds) in the 5 (blue) and 13 (red) electron regimes, plotted in Fig.~\ref{fig:plot_orbital_energies_1}(a). All the energy traces appear to be linear with the gate voltage. We also plot in Fig.~\ref{fig:plot_orbital_energies_1}(b) the extracted ellipticity, $\delta$, by taking the ratio between $\mathcal{E}_x$ and $\mathcal{E}_y$ in both the 5 and 13 electron configurations (blue and red respectively). We observe that $\delta$ varies linearly as a function of gate voltage $V_\mathrm{G2}$. In the dipolar coupling, the Rabi frequency given in Eq.(\ref{defRabiEDSR}) is proportional to $1/\sqrt{\omega_y}$ (present in the $y$ component of the dipole operator transition element). From Fig.~\ref{fig:plot_orbital_energies_1}(a), $\omega_y$ variation is very small, ruling out the possibility that the observed Rabi enhancement occurred because of non expected changes in $\omega_y$ when varying $V_\mathrm{G2}$.

\begin{figure}
        \centering
        \includegraphics[width=0.2\textwidth, angle = 0]{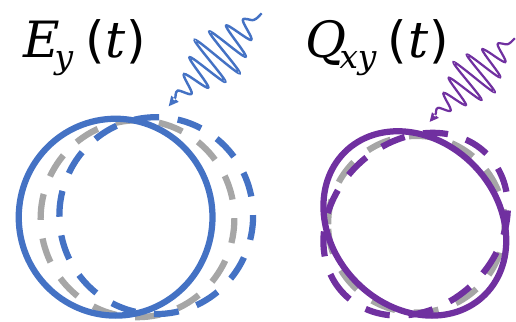}
        % \captionsetup{format=plain, font=small, labelfont=bf}
        \caption{\textbf{Cartoon of the effects of dipolar and quadrupolar electric fields.} 
       An AC quadrupole field results in a dilation/contraction of the electrostatic potential in orthogonal axes whereas an AC dipole electric field results in a shift.}
        \label{fig:Ey_and_Qxy}
    \end{figure}

    To account for these observations, our model should include the second order Taylor expansion of the driving electric field, or in other words, the quadrupole field
    \begin{equation} \label{eqn:Qxy_field}
        Q_{xy} = -\frac{\partial^2V}{\partial x \partial y} = \frac{\partial E_y}{\partial x}
    \end{equation}
    that couples to the qubit via the driving term $H_\mathrm{AC}^\mathrm{EQSR}(t)=|e|Q_{xy} xy\cos(\omega t)$. The full AC Hamiltonian is now extended as
    \begin{equation}
        H_\mathrm{AC}(t) = (|e|E_y y + |e|Q_{xy} xy)\cos(\omega t)
    \end{equation}
    which at resonance gives rise to what we call Electric Quadrupole Spin Resonance (EQSR), in analogy to Electric Dipole Spin Resonance. A schematic of how the quadrupole driving field affects the electrostatic potential can be found in Fig.~\ref{fig:Ey_and_Qxy}.

    The Rabi frequency will now be given by
    \begin{align}
        \hbar\Omega &= \left|\langle 0 | eE_y y + eQ_{xy}xy| 1 \rangle \right| \label{Rabi_eq:EQSR}\\ 
        & \propto |a_1\langle P_y,\uparrow | E_y y | S,\uparrow \rangle + a_0^*\langle S,\downarrow | E_y y | P_y,\downarrow \rangle + \nonumber \\
        & b_1\langle P_y,\uparrow | Q_{xy}xy | P_x,\uparrow \rangle + b_0^*\langle P_x,\downarrow | Q_{xy}xy | P_y,\downarrow \rangle +\cdots |\nonumber \;.
    \end{align}
    Here, we see that the Rabi frequency enhancement can also be promoted by the second-order perturbation terms of the states $|0\rangle$ and $|1\rangle$. In fact, as the energy difference between the $P_x$ and $P_y$ orbitals approaches that of the Zeeman splitting ($\hbar|\omega_x-\omega_y|\sim \mathcal{E}_\text{Zee}$)
    %{\color{blue}(that is, $|\hbar\omega_x-\hbar\omega_y|\sim \mathcal{E}_\text{Zee}$)}, 
    which can be achieved by tuning the gate voltages, the coefficient $|b_i|$ grows larger. Therefore, this treatment may predict the enhancement observed within this regime. For the 13-electron experiment, this quadrupole driving will also couple the $D_{xy}$ with $D_{yy}$ or $D_{xx}$ orbitals.
    
    Although quadrupole coupling is expected due to the non-uniformity of the electric field produced by the gates, it is often been neglected in calculations of the Rabi frequency. The $Q_{xy}$ term in the Hamiltonian is the only quadrupole component to directly couple the $P_x, P_y$ orbitals, as well as the $D_{xy}, D_{yy}$ ones, which can account for the observed Rabi enhancement. For this reason we neglect the other components.
    %There are other components of the quadrupole field in the driving term $H_\mathrm{AC}$, but they do not offer direct coupling between the orbitals and therefore become negligible in our description.

    %We conclude this section with some intuition on the influence of the quadrupole $Q_{xy}$ term on the dot.
    To conclude this section, whereas EDSR arises from a lateral oscillation of the dot, EQSR is caused by a dilation/contraction of potential along the $x=\pm y$ axes (see Fig. \ref{fig:Ey_and_Qxy}). Another way of thinking about this driving term is that it has an $x=y$ symmetry axis, which allows it to couple orbitals of opposite $x$ and $y$ symmetry. COMSOL simulations yield $Q_{xy}/E_y=1.2 \times 10^{-3}$ nm$^{-1}$, a ratio that we kept fixed when fitting our model to the experiment. Our dot occupies a region of linear size $L \sim 20$ nm, and therefore we see that $Q_{xy}L/E_y\ll 1$, as required for the validity of the multipolar expansion.

    \section{Results and discussion}
    \label{section:results_discussion}

\begin{figure*}[!ht]
        \includegraphics[width=0.95\textwidth, angle = 0, trim = 0cm 0cm 0cm 0cm]{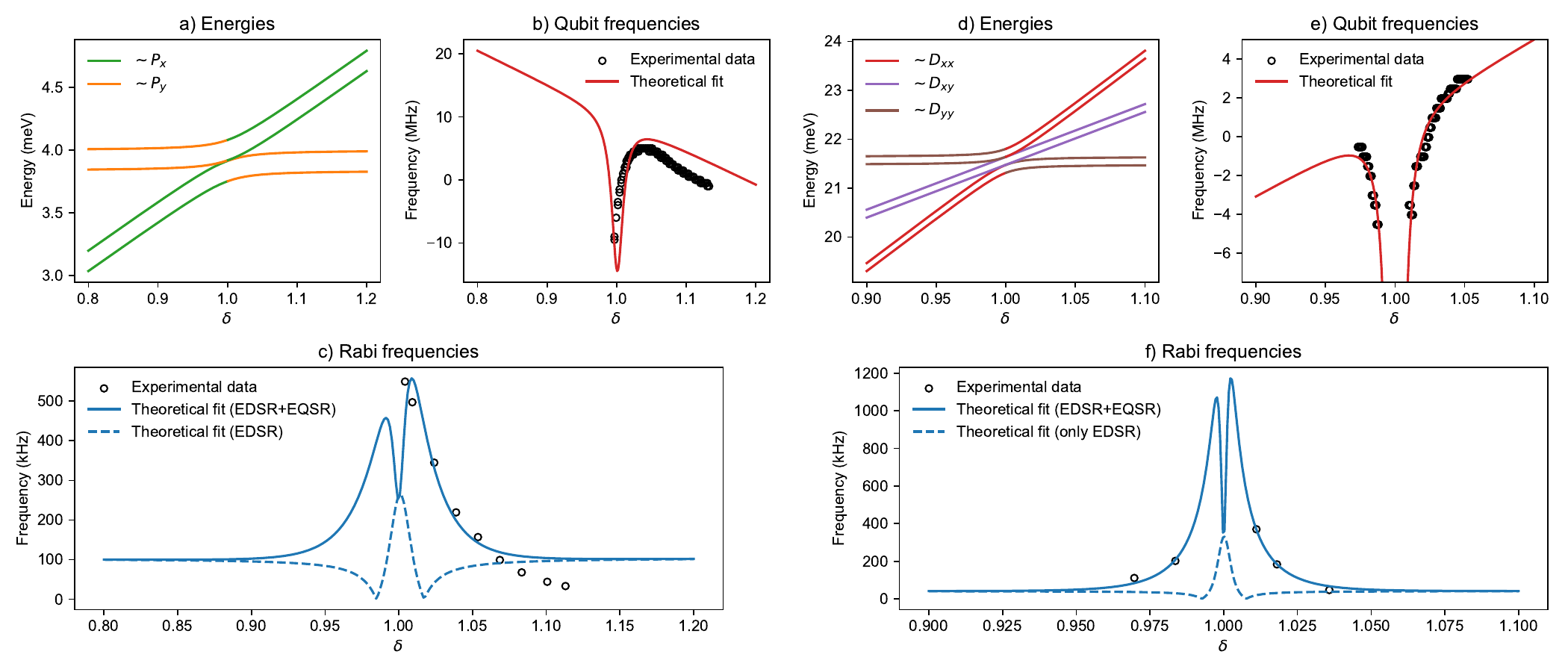}
        % \captionsetup{format=plain, font=small, labelfont=bf}
        \caption{\textbf{Fitted EQSR model.}
        (a) Energies of the $P_x$ and $P_y$ orbitals with the lowest two being the ones driven by EQSR.
        (b) Qubit frequency spectrum of the 5 electron regime, with the measured data in black and the fitted line in red.
        (c) Rabi frequency measured in the 5 electron regime, with the measured data in black and the fitted line in blue. The dashed line indicates the model prediction if only the electric dipole effect is accounted for.
        (d) Energies of the $D$ orbitals, with the lowest two being those driven by EQSR.
        (e) Qubit frequency spectrum of the 13 electron regime, with the measured data in black and the fitted line in red.
        (f) Rabi frequency measured in the 13-electron regime, with the measured data in black and fitted line in blue. The dashed line indicates the model prediction if only the electric dipole effect is accounted for. All $x$-axis labels here are the ellipticity $\delta \equiv \frac{\mathcal{E}_{x}}{\mathcal{E}_{y}}$.
        }
        \label{fig:fitted_EQSR}
    \end{figure*}

    In this section, we show that EQSR allows us to explain the experimental results reported by Ref. \cite{leon2020coherent}. Diagonalizing $H_\text{DC}$ [Eq.~(\ref{eq:h_dc})] allows us to obtain both the ESR frequency given by $hf_\text{ESR}=\mathcal{E}_1-\mathcal{E}_0$, and the Rabi frequency $\Omega$ given by Eq.~(\ref{Rabi_eq:EQSR}). We perform a least-squares fit to the experimental data, simultaneously fitting both the qubit and Rabi frequencies as a function of the ellipticity of the dot $\delta$, with results shown in Fig.~\ref{fig:fitted_EQSR}. A similar fitting protocol was used in Ref.~\cite{gilbert2023demand}.

    We first examine the results of the 5-electron regime, with the resulting energy levels shown in Fig.~\ref{fig:fitted_EQSR}(a). Here, we are largely concerned with the $P_x$ and $P_y$ orbitals, given that (for $\delta>1$) we expect the valence electron to be sitting in the $P_y$ orbital based on electron-filling principles (and in the $P_x$ orbital for $\delta<1$). We can obtain the qubit frequencies by calculating the energy difference between the ground and the first excited state, which we plot in Fig.~\ref{fig:fitted_EQSR}(b). Here, the theoretical estimate is presented in red, contrasting with the experimental data in black. Finally, we can calculate the resulting Rabi frequency based on these qubit parameters, which is plotted in Fig.~\ref{fig:fitted_EQSR}(c). 
    %{\color{blue}The solid blue line indicate the Rabi frequency found accounting EDSR and EQSR, while the dashed blue line shows what's the contribution of EDSR for this result.}  
    The solid blue line indicate the Rabi frequency found accounting EDSR and EQSR, while the dashed blue line singles out the EDSR contribution.
    We also tried to do a fitting procedure that does not include EQSR, but we could not reproduce the enhancement that follows the experimental data near the degeneracy point. This supports our theoretical hypothesis that EQSR is required to explain the Rabi frequency speed-ups. Physically this is due to the fact that with EDSR alone the Rabi frequency would present a maximum when there is an anticrossing in Fig. \ref{fig:fitted_EQSR}(a). Nonetheless, we observe a maximum where the Stark shift (given by the slope in Fig. \ref{fig:fitted_EQSR}(b) is maximum, a feature only accounted by the EQSR coupling, as we can see in Fig. \ref{fig:fitted_EQSR}(c).
    Furthermore, the peak in the dashed lines of Figs \ref{fig:fitted_EQSR}(c) and (f) can be shown to be a higher order effect \cite{supp}.
    
   % \sout{, with the fit being significantly better with} 
    %\sout{Taking the difference in energy between the $P_y$ orbitals, we can obtain the qubit frequency shown in Fig.~\ref{fig:fitted_EQSR}(b), where we plotted the theoretical estimation in blue against the experimental data in purple. Finally, we observe the resulting speed-up in Rabi frequency in Fig.~\ref{fig:fitted_EQSR}(c), where we show that the theoretical model predicts the speed-up well, which would not be captured in the case of only having \blue{electric dipole} contributions.}

    We obtained similar results in the 13-electron regime, with the main difference being that the electric drive is acting on the $D$ orbitals. We show the $D$ orbital energy levels, the qubit frequency fits, and the Rabi frequency fits in Figs.~\ref{fig:fitted_EQSR}(d), (e), and (f) respectively. The qubit frequency is similarly defined as the difference between the ground and first excited energy of the $D$ orbital energy levels. In Fig.~\ref{fig:fitted_EQSR}(f), we show that the enhancement in Rabi frequency can be mapped by the Hamiltonian with the inclusion of EQSR. 
    
    %In both cases, we show that we are able to accurately capture the behavior of the Rabi frequency speed up, even though the fitting of the qubit frequency is non-ideal. These fittings are difficult for multiple reasons, including but not limited to the lack of sufficient experimental data points and the numerous parameters. We will discuss the technical aspects of the fitting protocol in greater detail in the Supplementary Material \cite{supp}.
    
    Therefore, in both the 5 electron and 13 electron cases, we demonstrate our ability to capture the behavior of the Rabi frequency speed-up.  We stress that a quantitative correspondence between the experimental data and the theoretical fits is not the scope of this paper, but we show how the electrical quadrupole is necessary to explain the Rabi enhancement near the degeneracy point. The challenges in these fittings arise from several factors, including, but not limited to, insufficient experimental data points and having a multitude of parameters. 
    %We will discuss the technical aspects of the fitting protocol in greater detail in the Supplementary Material.

    From a physics point of view, we know that the ESR frequency of the qubit can be susceptible to the surface roughness at the silicon-silicon dioxide interface \cite{cifuentes2024bounds} as well as nuclear spins \cite{stano2023dynamical}, and that is difficult to account for in fittings such as the one performed here. We would have to include some form of non-linear voltage dependence in the qubit frequency.
    %The fits show the comparison of the qubit and Rabi frequency between the experimental data and the theoretical calculations with and without the effects of EQSR for experiments with 5 and 13 electrons; and the energy of the unperturbed states to highlight the anticrossing region. See table \ref{tab_fit_values} for the fitted values.

    Finally, we see from Eq.~(\ref{Rabi_eq:EQSR}) that while Rabi frequency promoted by EDSR depends on $a_{0(1)}\propto\alpha+\beta$, the Rabi enhancement promoted by EQSR depends on $b_{0(1)}\propto\alpha^2-\beta^2$, which is key relation to consider in order to improve Rabi frequencies next to the degeneracy point using intrinsic spin-orbit coupling. If the difference between the square of the two terms are large enough, the addition of a micromagnet could become unnecessary, which is desirable for scale-up architectures. The EQSR theory not only explains behavior in previous experiments, but is also a formalism that accounts for phenomenon in the electrical control of quantum dots qubits.

    The experimental results in Ref.~\cite{leon2020coherent} show a clear correlation between the nonlinear bend of the qubit frequencies and the one order of magnitude increase in the Rabi frequencies of $P$ and $D$ orbitals. Both features can be explained within an effective single-particle theory by taking quadrupole coupling into account. Our fits show the rising of Rabi frequency as dot shape becomes more circular, i.e. as we approach the anti-crossing point, therefore indicating clearly the role of EQSR in enhancing the Rabi frequency close to the degeneracy as a function of the ellipticity. 
    
    This work opens the perspective of analyzing how the EQSR combined with other coupling affects the driving mechanism, for example with valley-orbit coupling, which has recently been shown to perform EDSR\cite{Corna2018}. Another perspective is to analyze the influence of electron-electron interaction on the EQSR mechanism. In summary, we believe that the technique of quadrupole spin resonance opens up new pathways for coherent spin control with multi-electron dots and can become an ally in the quest for scalable quantum computation.

\section{Acknowledgments}
We acknowledge support from the Australian Research Council (FL190100167, CE170100012, and IM230100396) and the US Army Research Office (W911NF-23-10092). The views and conclusions contained in this document are those of the authors and should not be interpreted as representing the official policies, either expressed or implied, of the Army Research Office or the US Government. M.K.F. and J.D.C acknowledges support from Sydney Quantum Academy. L.A.L. and P.H.P. acknowledge funding from the Brazilian agencies Conselho Nacional de Desenvolvimento Cient\'ifico e Tecnol\'ogico (CNPq) and Coordena\c c\~ao de Aperfei\c coamento de Pessoal de N\'ivel Superior (CAPES).

\bibliography{pubs}
\end{document}